# Q-DPM: An Efficient Model-Free Dynamic Power Management Technique

Min Li[1,2], Xiaobo Wu[1], Richard Yao[2], Xiaolang Yan[1]

[1] *Institute of VLSI Design, Zhejiang Univ., China*    [2]*Microsoft Research Asia*

## Abstract

When applying Dynamic Power Management (DPM) technique to pervasively deployed embedded systems, the technique needs to be very efficient so that it is feasible to implement the technique on low end processor and tight-budget memory. Furthermore, it should have the capability to track time varying behavior rapidly because the time varying is an inherent characteristic of real world system. Existing methods, which are usually model-based, may not satisfy the aforementioned requirements. In this paper, we propose a model-free DPM technique based on Q-Learning. Q-DPM is much more efficient because it removes the overhead of parameter estimator and mode-switch controller. Furthermore, its policy optimization is performed via consecutive online trialing, which also leads to very rapid response to time varying behavior.

## 1. Introduction

Dynamic Power Management (DPM) is a very general and important power reduction technique targeting on controlling performance and power levels of electronic systems. In DPM system, a Power Manager (PM) monitors the overall system states, and controls the power state of system components. In the coming deep submicron era, DPM will gain more and more importance, because DPM is the one of the most effective techniques to combat with leakage current.

Actually DPM has attracted a lot of interests in the past a few years [1]. Most of them apply stochastic modeling and optimization. The existing methods usually assume that DPM is controlling a stochastic system with explicit model completely known. Recently, some papers have studied how to handle the uncertainty and nonstationary characteristics of the real world. However, these techniques are still based on model.

In the near future, a lot of research challenges will be imposed on DPM. Above all, DPM is demanded by deeply embedded and pervasively employed smart nodes around us, e.g., biosensor node. They have only low end processor and tight budget memory. Hence, the DPM policy optimization needs to be extremely efficient. Actually the same requirement exists in large system and System on Chip as well. Moreover, as discussed in some previous papers, in most real world systems parameters are undertaking continuous varying, and the varying behavior needs to be rapidly tracked, so that the maximum potential of power reduction can be delivered.

Existing techniques have a number of shortcomings that prevent them from surviving the research challenges. First of all, the existing model-based techniques are impractical for low-end deeply embedded systems because of the complexity of policy optimization and time penalty for dealing with time varying. In experiments we observe that, even on Pentium III 800MHz PC, the widely applied linear programming policy optimization runs extremely slow. Besides policy optimization, the parameter estimation also consumes a lot of time to maintain a reasonable accuracy. In addition, the model-switch controller, which detects parameter variation and determines when to perform policy optimization, is fairly time consuming as well. Hence, when parameters vary continuously, existing model-based approaches may not track the varying rapidly.

In this paper, we present a Reinforcement Learning (RL) based DPM technique namely Q-DPM. The proposed technique is model free, i.e., it does not require explicit model of the system. On the contrary, it learns the optimal policy by trialing continuously. The complexity of the technique is much lower than the model-based counterpart, so that it is feasible to implement on almost any low end systems. Furthermore, since Q-DPM performs consecutive adaptation in every time interval and it removes the overhead of parameter estimator as well as mode-switch controller, its response to varying behavior is very rapid.

## 2. Q-DPM

RL is the problem faced by an agent that must learn behavior though trial and error interactions with a dynamic environment. In the standard RL mode, an agent is connected to its environment via perception and action. On each step of interaction the agent receives input as some indication of the state of the environment. The agent then chooses an action, which changes the state of the environment. The value of this state transition is communicated to the agent through a reinforcement signal. The agent's behavior should choose actions that tend to increase the long-run sum of values of the reinforcement signal. It can learn to do this over time by systematic trial and error.

From above description we can see that the RL is one way to derive the policy for Discrete Time Markov

Decision Process (DTMDP). RL methods solve DTMDP by learning good approximations to the optimal value function, $J^*$, given by the solution to the Bellman optimality equation [3] which takes the following form

$$J^*(s) = \max_{a \in A(s)}(E_{s'}(c(s,a,s') + \beta J^*(s'))), \quad (1)$$

where $A(s)$ is the available action set in the current state $s$, $c(s,a,s')$ is the effective immediate payoff (in the context of DPM. The payoff is energy reduction or certain function of energy reduction), and $E_{s'}$ is the expectation over possible next states $s'$, and $\beta$ is the discount factor. Also, RL can be interpreted as direct adaptive control [4].

We adopt Q-Learning [5] for our model free DPM. Q-Learning is almost the most practical RL algorithm because it is quite easy to implement. When applying the Watkin's Q-Learning, Bellman's equation can be rewritten in Q-factor as

$$J^*(s) = \max_{a \in A(s)} Q^*(s,a), \quad (2)$$

where $Q^*(s,a)$ can be interpreted as the expected discounted reinforcement (actually a function of energy reduction) of taking action $a$ in state $s$. Intuitively, on each step we can greedily take the action with maximum $Q(s,a)$. On a transition from state $s$ to $s'$ with action $a$, the $Q(s,a)$ is updated as following:

$$Q(s,a) = (1-\gamma)Q(s,a) + \gamma(c(s,a,s') + \lambda \max_{b \in A(s')} Q(s',b)), \quad (3)$$

where $\gamma$ is the learning rate. In order to make Q-Learning perform well, all potentially important state-action pairs $(s,a)$ must be explored. At each state, with probability $\chi$ a random action needs to be taken instead of the action recommended by the $Q(s,a)$.

Apparently the run time complexity of Q-DPM is very low. On each step, the DPM daemon only needs to select the maximum $Q(s,a)$, and update the $Q(s,a)$ using Eqn. 3. $Q$ values can be encoded in a $|s| \times |a|$ table that requires a little bit memory space. Hence, it is feasible to implement Q-DPM on almost any embedded nodes. Actually, Q-DPM has a number of additional advantages, and the most attractive one is that it is tolerant to small scale variations, which is very likely to happen in real world system.

## 3. Simulation and Conclusion

The simulation is to preliminarily investigate the performance of Q-DPM. In order to study Q-DPM under various conditions, synthetic input is used to drive the simulation. First we study whether Q-DPM can deliver the maximum potential of DPM in stationary environment. In Fig.1 we compare Q-DPM to the optimal policy derived by analytical techniques which assume model is completely known in prior. After studying many cases, we conclude that Q-DPM can approximate the theoretically optimal policy at reasonable speed despite it requires much less resources. Moreover, we put Q-DPM in a nonstationary environment by feeding temporarily stationary synthetic input. One typical case is shown in Fig. 2, where switching points are marked by vertical lines. We can read that the energy reduction may be heavily affected by parameter variation (e.g., around the first changing point), and the proposed Q-DPM responds to the variations almost instantly. In contrast to Q-DPM that directly learns optimal state-action mapping, existing methods need to detect parameter change, perform parameters, and then perform time consuming policy optimization. The significant time overhead is removed in Q-DPM.

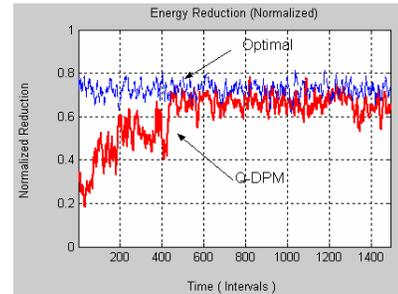

**Fig. 1 Convergence on Optimal Policy**

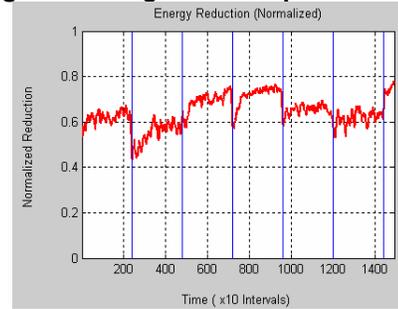

**Fig. 2 Rapid Response**

There is still a lot of rewarding research remaining to perform, such as QoS guaranteed Q-DPM and Fuzzy Q-DPM in noisy environment. More importantly, the proposed approach is to be verified in experiments.

## 4. Acknowledgement

This work is supported by the National Natural Science Foundation of China under grant No. 90207001.